\documentstyle[12pt,psfig]{article}

\begin{document}
\thispagestyle{empty}
\baselineskip=24pt
\thispagestyle{empty}
\newcommand{\etal}{{\it etal.\/}}

\begin{center}
	{\large \bf  Electron paramagnetic resonance studies of the insulating ferromagnetic manganite Nd$_{0.8}$Pb$_{0.2}$MnO$_{3}$ above the transition temperature} %T$_{C}$\\}
	\normalsize
\end{center}

\begin{center}
	{\bf Anika Kumar\footnote {Indian Academy of Sciences summer fellow; Present address: Department of Electrical, Computer and Systems Engineering, Rensselaer Polytechnic Institute, Troy, New York 12180}, Nilotpal Ghosh, Janhavi P. Joshi, H.L. Bhat \\and S.V. Bhat\footnote {Correspnding author. Tel.:+91-80-3942727;fax:+91-80-3602602\\
{\it E-mail address:} svbhat@physics.iisc.ernet.in}\\}
	
	{ Department of Physics, Indian Institute of Science, Bangalore-560012,India\\}
	\end{center}

\thispagestyle{empty}
\baselineskip=18pt
\vskip.02cm

\begin{abstract}
Single crystals of Nd$_{1-x}$Pb$_{x}$MnO$_{3}$ with x=0.2 are grown by high temperature solution growth technique using PbO-PbF$_{2}$ flux. Magnetization studies on the samples show a transition to a ferromagnetic state below T$_c$ $\sim$ 125 K and the resistivity measurements show it to be an insulator throughout the temperature range 50 - 300 K. Electron Paramagnetic Resonance studies have been performed for T $> T_{C}$ with a view to comparing the results with those on metallic ferromagnetic manganites. The temperature dependence of various parameters like g-value, linewidth and intensity has been studied in the temperature range 150  - 300 K. It is found that they behave in a manner similar to that exhibited by metallic ferromagnetic manganites.  
\vskip0.5cm
\noindent {\it PACS}: 76.30.-v, 75.70.Pa, 72.80.Ga, 71.30.+h
\vskip0.5cm
\noindent {\it Keywords}:A:Insulators A: magnetically ordered materials B: Crystal growth E: Electron Paramagnetic Resoanance

\end{abstract}

\newpage

{\large{\bf 1. Introduction :}\\}

Hole-doped rare earth manganites of the form  R$_{1-x}$B$_{x}$MnO$_{3}$ where R is a trivalent rare earth ion such as  La$^{3+}$, Pr$^{3+}$, Nd$^{3+}$ etc. and B is a divalent alkaline earth ion like Ca$^{2+}$, Sr$^{2+}$, Ba$^{2+}$ and Pb$^{2+}$ have generated intense scientific activity in recent years because of the very interesting structural, transport and magnetic properties they exhibit as a function of composition and temperature\cite{rao,tok}. For example, when cooled below a certain temperature, most of these compounds undergo an insulator to metal transition which is usually accompanied by a paramagnetic to ferromagnetic transition as well around the same temperature. More interestingly they also show a very large negative magnetoresistance, aptly termed  Colossal Magneto-Resistance (CMR), around this temperature. CMR is understood in the framework of the Zener double exchange (DE) model which is based on strong Hund's coupling existing between the mobile e$_g$ electron and the core t$_{2g}$ electrons of the neighboring Mn$^{3+}$ and Mn$^{4+}$ ions. The charge transport is understood to retain, indeed require, parallel orientation of the e$_g$  and the t$_{2g}$ electrons thus resulting in an intimate connection between the metallic conductivity and the ferromagnetic order. 

Recently however, it has been realised that DE model may not be able to fully describe certain properties of manganites. For example, in the low doping regime, the system, even in the ferromagnetic state, is found to be an insulator, thus apparently contradicting the DE model. There also has been evidence for the mixed ferromagnetic insulating (FMI) and ferromagnetic conducting (FMC) phases in certain regions of the x-T phase diagram\cite{dag}. While some theoretical models propose that such mixed phases are a consequence of electronic phase separation, experimentally it is still not known whether the mixed phases have the same electronic properties as the constituent pure phases or if they have entirely different properties of a ``novel" electronic state. 

Electron Paramagnetic Resonance (EPR) is expected to be a very useful technique to study manganites since it is a sensitive probe of spin states and their dynamics. A number of EPR studies of mostly metallic ferromagnetic manganites in their paramagnetic state has been published in the last few years focusing on the spin dynamics vis-a-vis the charge transport\nocite{ose,riv,tov,cau,iva,she,she1,lof}[4-11]. In this work we present the results of our EPR experients on an insulating ferromagnet Nd$_{0.8}$Pb$_{0.2}$MnO$_{3}$, in its paramagnetic state,  carried out with a view to comparing them with those of the metallic ferromagnets.
\newpage
{\large{\bf  2. Experimental Details :}\\}

Single crystals of Nd$_{1-x}$Pb$_{x}$MnO$_{3}$ with x=0.2 are prepared by the high temperature solution growth technique \cite{morr} using a PbO-PbF$_{2}$ flux . For synthesis, the precursors, Nd$_{2}$O$_{3}$, MnCO$_{3}$, PbO and PbF$_{2}$ were weighed in appropriate quantities and homogenized in a ball mill. The homogenized mixture was then subjected to heat treatment under a controlled temperature environment facilitating single crystal growth.

The dc magnetization was measured using a commercial vibrating sample magnetometer(VSM) at 5 kOe magnetic field in the temperature range 280 to 4.2K. The resistivity measurement done using four probe method showed an insulating phase from room temperature down to 50 K \cite{Nilo}. The EPR measurements were carried out at 9.4 GHz with a Bruker ER-200D-SRC spectrometer on single crystals as well as on powder samples obtained by crushing the single crystals. The powder was dispersed in paraffin wax and DPPH was used as a `g' marker. The spectrometer was equipped with a 12-bit A to D converter interfaced to a personal computer. In the typical field scan of 6000 G, this arrangement gives a precision of $\pm$ 3 G. The temperature was varied between 4.2 - 290 K using an Oxford instruments continuous flow cryostat and temperature regulator (accuracy $\pm$ 1 K).  We present here results only in the temperature range of 150  - 300 K, i.e. in the paramagnetic state of the material. The EPR results in the ferromagnetic state of the manganites are known to be quite complex being dependent on sample size and thermal cycling in addition to the influence of the local field. 
\vskip1cm

{\large{\bf 3. Results and Discussion :}\\}

The signals from the single crystal have  Dysonian lineshapes at room temperature (even though the material is an insulator, it has a moderate conductivity of 1.9 Siemens~cm$^{-1}$ at room temperature) and  they become very complex and distorted due to the effects of demagnetization factors and internal fields as the ferromagnetic order begins to set in at low temperatures. Hence, in this report we concentrate on the data from the powder sample. 

In fig. 1 we show typical EPR signals obtained from the powder sample of 
Nd$_{0.8}$Pb$_{0.2}$MnO$_{3}$. The signals are broad with symmetric lineshapes which could be fitted to the derivative of the Lorentzian function given by,

\begin{equation}
\frac{dP}{dH}\propto \frac{d}{dH}\left(\frac{\Delta H}{4(H-H_{0})^{2}+\Delta H^{2}}+\frac{\Delta H}{4(H+H_{0})^{2} + \Delta H^{2}}\right)
\end{equation}

where $\Delta$H is the full width at half maximum and  H$_0$ is the resonance field. The signal being broad we have used the two terms in the equation representing the response of the system to the microwave components polarised clockwise and anticlockwise respectively \cite{iva}. 

The lineshape parameters obtained by fitting the above equation to the EPR signals in the temperature range of 150 - 290 K are plotted as  functions of temperature in fig 2. Fig. 2a shows the peak-to-peak linewidth $\Delta$H$_{pp}$ plotted against temperature ($\Delta$H$_{pp}$= $\Delta$H$_{FWHM}$/$\sqrt{3}$). As a function of T, it showed a behaviour similar to that described in previous papers on ferromagnetic manganites in their paramagnetic states: on approaching the T$_{C}$ from above, $\Delta H_{pp}(T)$  goes through a minimum at \(T_{min}>T_{C}\)(fig. 3). Above T$_{min}\approx 149K$, the variation of $\Delta$H$_{pp}$(T) is found to be approximately linear[4-11].
In fig. 2b we have plotted g as a function of temperature. The g value also shows a monotonic decrease as a function of temperature from 150  to 290 K.
%The intensity of the signal I(T) is determined by numerical double integration of the measured spectra and is shown 
%in fig. 2c. The inset to fig. 2c shows the dc susceptibility $\chi_{dc}$ as a function of temperature. As can be seen, the intensity goes on decreasing with increasing temperature and follows the dc susceptibility behaviour as a function of temperature from 150 K to 290 K.  

In figure 3 we have plotted the inverse EPR intensity as a function of temperature. We found that the inverse of EPR intensity, fitted the ferromagnetic (FM) Curie-Weiss behaviour with  $\Theta$= 148 K. As shown in the inset of fig. 3, at high temperatures, we observed that $\chi_{dc}$ also follows the Curie - Weiss temperature dependence, $\chi_{dc}(T)=C/(T-\Theta)$. From the linear behaviour of $\chi^{-1}_{dc}(T)$, we have determined the Curie constant as 0.021 emu K gm$^{-1}$ Oe$^{-1}$ and $\Theta$ is found to be 152~K which are very similar to the ones obtained from the $I^{-1}_{EPR}$(T). 

  Previously, Causa et al.,\cite{cau} and  Moreno et al.,\cite{mor} have interpreted the temperature  dependence of $\Delta H_{pp}$ of CMR manganites (which show ferromagnetic metallic phase),  in their  paramagnetic (PM) phase,  using the expression,  
\begin{equation}
	\Delta H_{pp}(T)\propto[\chi_{s}(T)/\chi_{epr}(T)]\Delta H_{pp}(\infty)
\end{equation}
where $\chi_{s}(T)=C/T$ is the single ion susceptibility and the $\chi_{EPR}(T)$ corresponds to the paramagnetic behaviour of the coupled system. Causa et al., also found that the ratio $\Delta H_{pp}(T)$/$\Delta H_{pp}(\infty)$ plotted as a function of $T\over{T_c}$ follows a universal curve. We have done the analysis of the EPR linewidth in Nd$_{0.8}$Pb$_{0.2}$MnO$_3$ in a similar fashion.
 Using the constants C and $\Theta$ obtained from the Curie-Weiss fit of $\chi^{-1}_{d.c}(T)$, we have plotted the peak to peak linewidth against (T-$\Theta$)/T (Inset of fig. 4). The linewidth shows a linear behaviour and  the slope of the linear fit gives $\Delta$H($\infty$). As can be seen from figure 4 the ratio of  $\Delta$H$_{pp}$(T) to
$\Delta$H$_{pp}$($\infty$) (multiplied by a scaling factor of 1.3 which incorporates a proportionality constant between I$_{EPR}$ and $\chi_{EPR}$ inclusive of filling factor etc.) plotted against the normalised temperature t=T/T$_c$ follows the universal curve  obtained by Causa et al., in case of CMR manganites. Thus, though the system under consideration undergoes a transition to a FMI phase, in the paramagnetic phase it shows essentially a similar spin dynamics as the CMR manganites studied by Causa et al.,

Finally, the g value (fig. 2b), which goes on increasing as the sample is cooled towards the FM transition temperature T$_c$, seems to suggest a building up of FM correlations. As the internal fields due to the FM correlations increase, the resonant field shifts downwards resulting in a monotonic increase in the g value with decreasing temperature. Previous EPR studies performed on CMR manganites do not report any in-depth study of the temperature dependence of g in the PM state although the FM correlations are expected to build up even in those systems. 

Another possible origin of this behaviour of g could be in the different electronic structure in the two types of systems. It has been found that in the FM metallic phase, the macroscopic Jahn-Teller ordering and distortion of octahedra disappear and ordinarily no orbital ordering is expected in this phase. However, experimental data suggest a possibility of orbital order in the FMI state. Indeed orbital order is considered to be a possible candidate for explaining the FMI state in the low doping regime of the phase diagram of manganites\cite{kho}. In the charge ordered compounds which we have studied before\cite{raj,jan}, we have attributed the increase in the g value below the charge ordering temperature, to the gradual build up of orbital order, resulting in the increase  in the spin-orbit coupling constant $\lambda$. The g shift observed in the manganites showing FMI phase, may also be attributed to the building up of such an orbital order. However, to substantiate this hypothesis, more experimental and/or theoretical work is required. 

{\large {\bf Acknowledgements}}

A.K. thanks the Indian Academy of Sciences, Bangalore for the award of a summer fellowship. The authors thank DST and CSIR, Govt. of India for financial support through project grants. J.P.J. thanks CSIR, India for the senior research fellowship.

\newpage
\begin{large}
\noindent {Figure Captions}\\
\end{large}
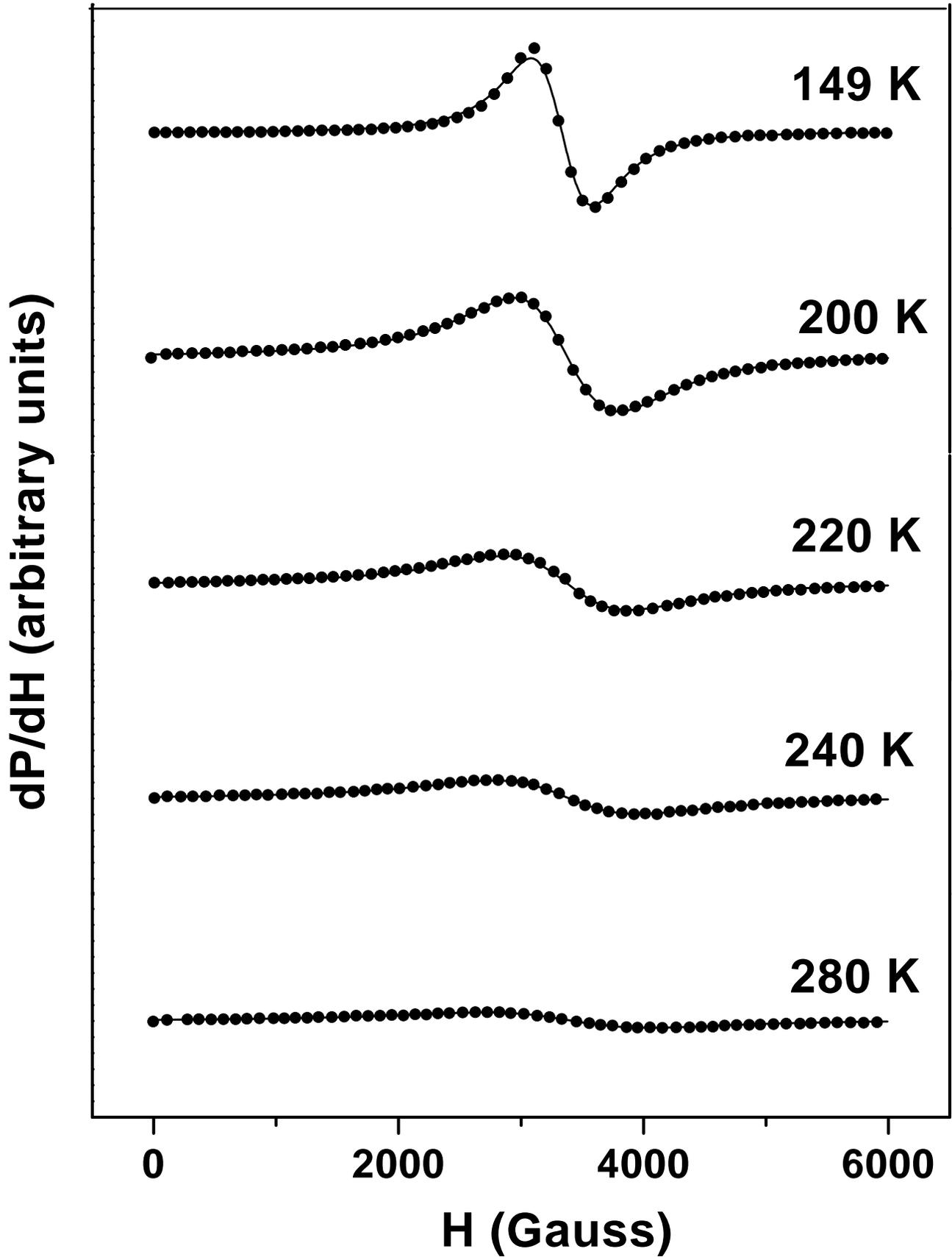
\noindent { FIGURE 1}

  EPR spectra of a powder sample of Nd$_{0.8}$Pb$_{0.2}$MnO$_3$ for a few representative temperatures. The solid circles are experimental data and the solid line shows the fit to  equation 1 described in the text. \\ 

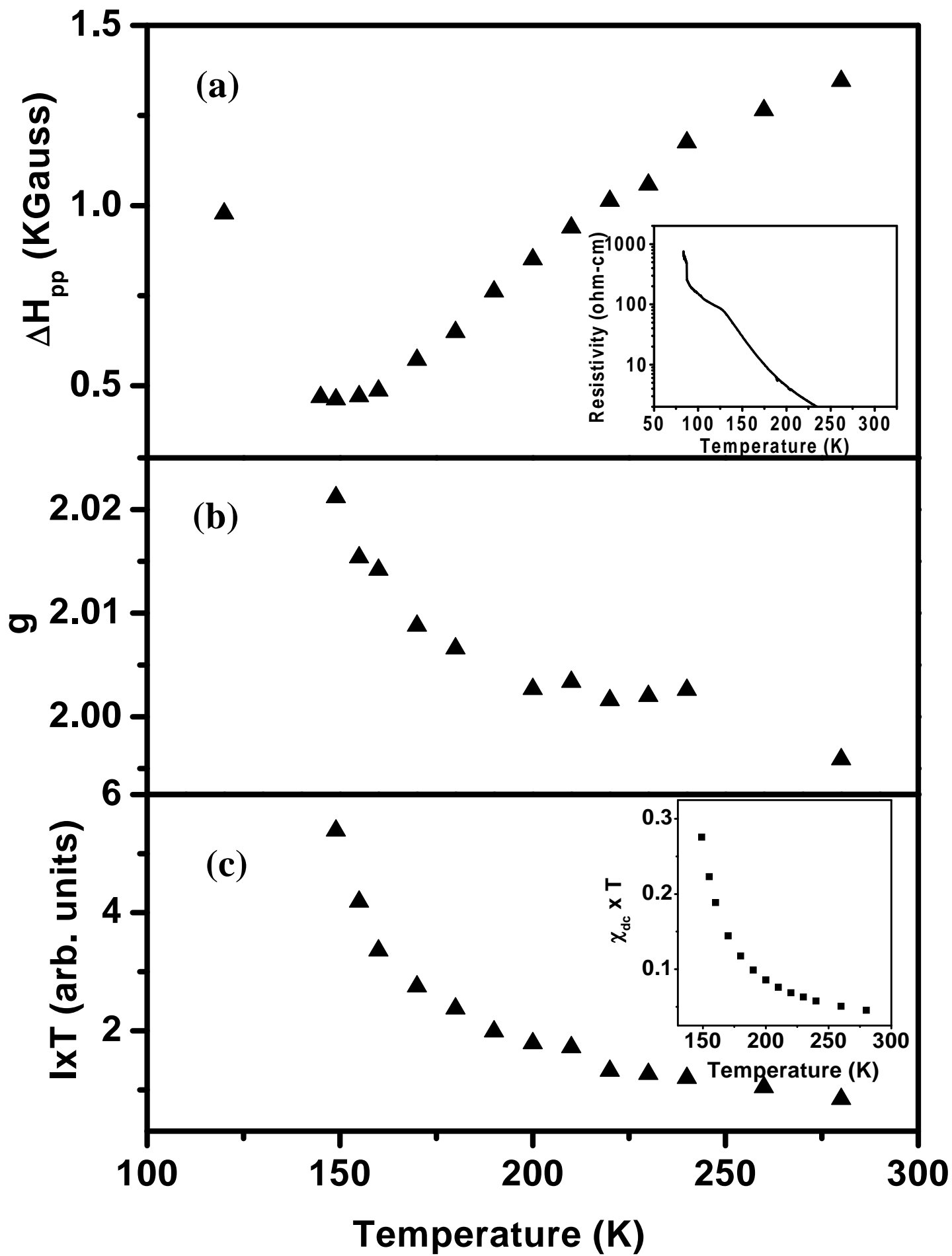
\noindent {FIGURE 2}

Temperature dependence of powder lineshape parameters obtained from the fits to equation 1;  (a) peak to peak linewidths $\Delta$H$_{pp}$,    (b)g value and (c)intensity times T. The inset to (a) shows resistivity {\it vs} T and that to (c) shows $\chi_{d.c}$ {\it vs} T. \\

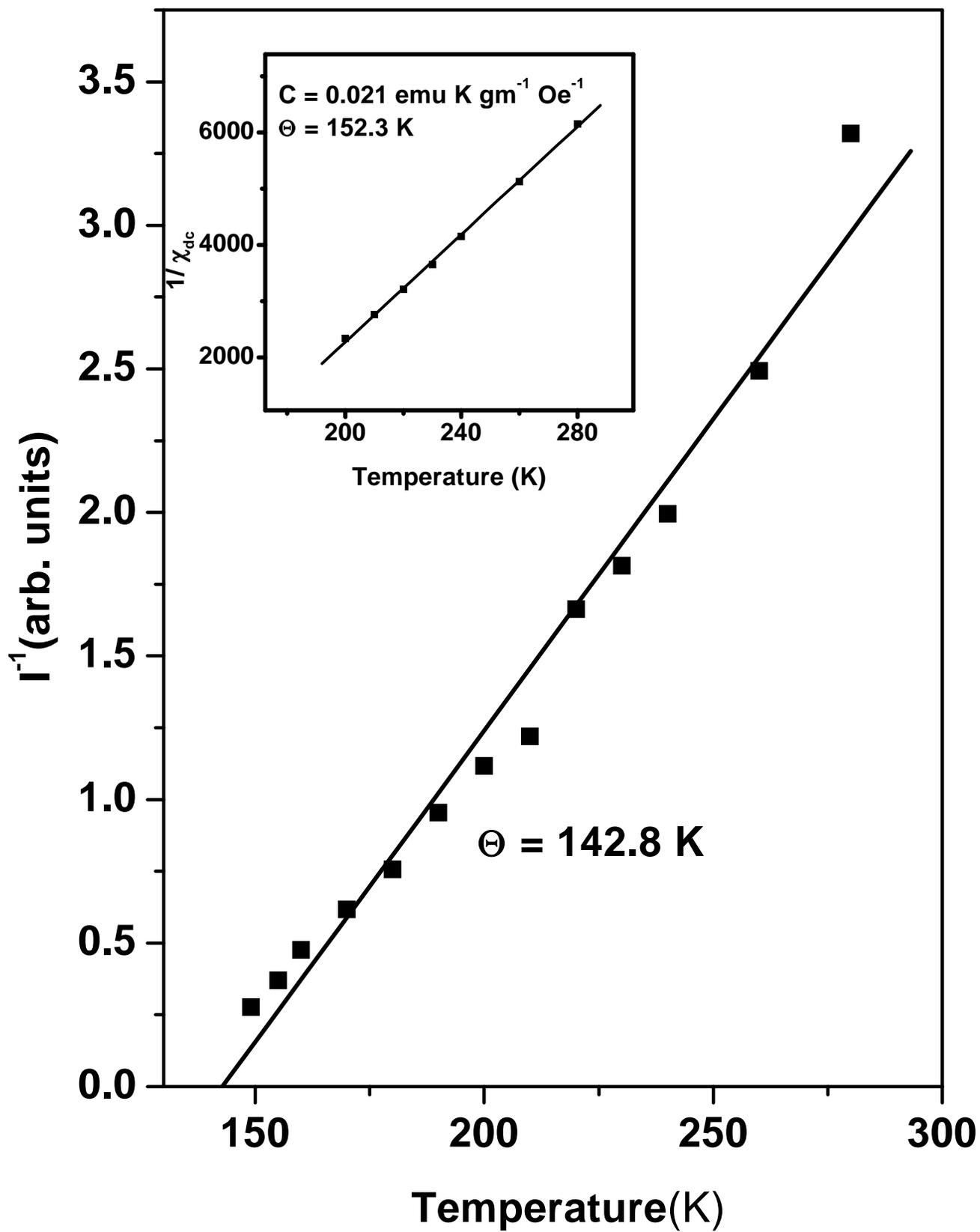
\noindent {FIGURE 3}

Temperature dependence of  I$^{-1}_{EPR}$; the solid squares are the experimental data and the solid line is a fit to the ferromagnetic Curie-Weiss behaviour. The inset shows 
 $\chi^{-1}_{d.c}(T)$  fitted to the Curie-Weiss law. The solid squares are experimental data and the solid line the fit.\\

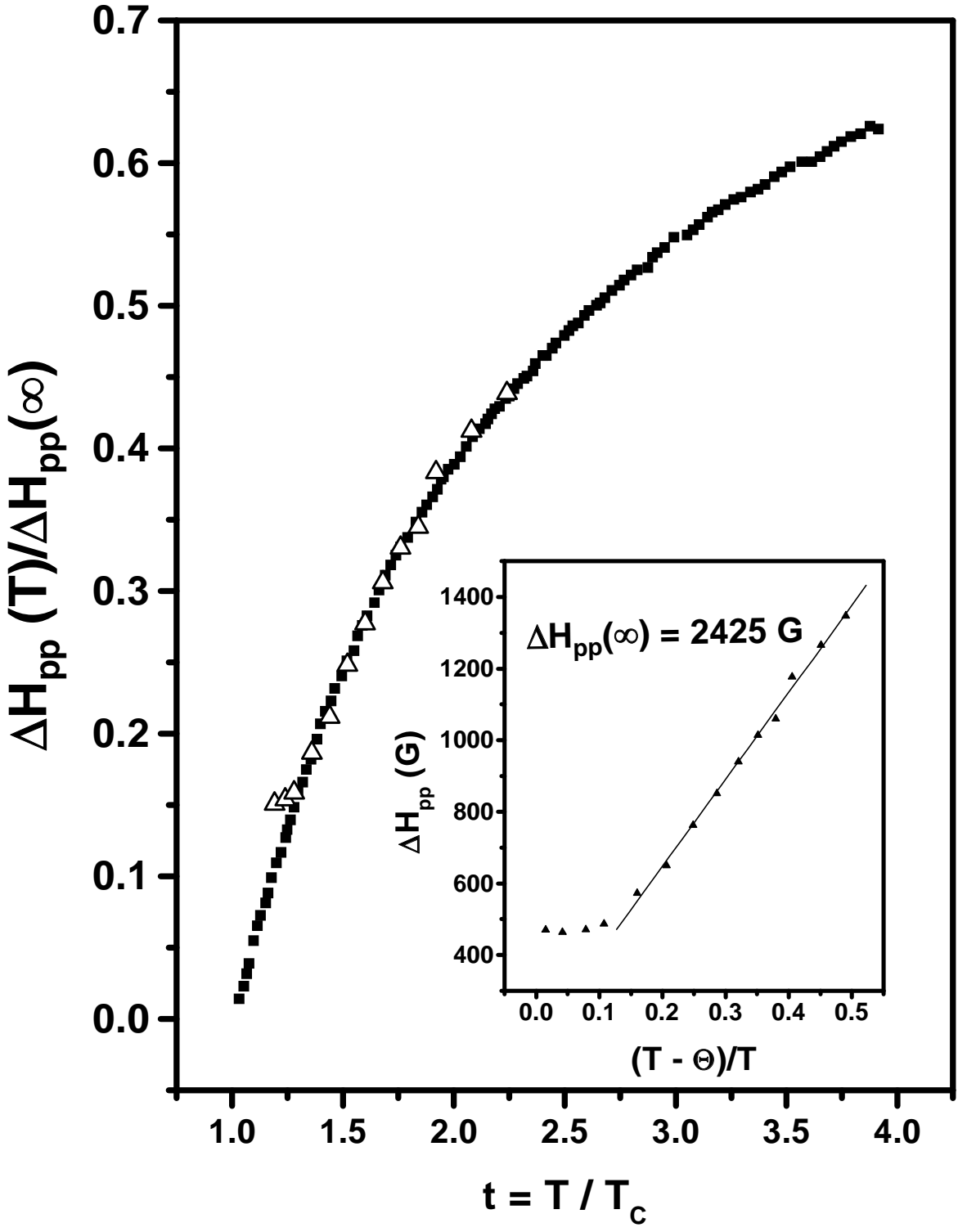
\noindent {FIGURE 4}

$\Delta$H$_{pp}$(T)/$\Delta$H$_{pp}$($\infty$) (open traingles) plotted as a function of T/T$_c$. The solid squares are the universal curve obtained from reference 6. The inset shows $\Delta$H$_{pp}$ plotted against (T-$\Theta$)/T. Solid line is a linear fit to the experimental data (solid traingles). Slope of the fit gives $\Delta$H$_{pp}$($\infty$).

\end{document}